\documentclass[%
 reprint,
superscriptaddress,
nofootinbib,
 amsmath,amssymb,
 aps,
prl,
]{revtex4-1}

\usepackage{graphicx}% Include figure files
\usepackage{subcaption}
\usepackage{dcolumn}% Align table columns on decimal point
\usepackage{bm}% bold math
\usepackage{siunitx}% non italicized mu for micro
\usepackage{color}
\definecolor{red}{rgb}{1,0.,0}
\definecolor{white}{rgb}{1.,1.,1.}

\begin{document}

%\preprint{APS/123-QED}

\title{Truncated nonlinear interferometry for quantum enhanced atomic force microscopy}% Force line breaks with \\

\author{R.C. Pooser}
\affiliation{Oak Ridge National Laboratory, Oak Ridge, Tennessee 37831, USA}
\author{N. Savino}
\affiliation{Oak Ridge National Laboratory, Oak Ridge, Tennessee 37831, USA}
\affiliation{Department of Physics and Engineering Physics, Tulane University, New Orleans, Louisiana 70118, USA}
\author{E. Batson}
\affiliation{Oak Ridge National Laboratory, Oak Ridge, Tennessee 37831, USA}
\affiliation{Department of Physics, Massachusetts Institute of Technology, Cambridge, Massachusetts 02139, USA}
\author{J.L. Beckey}%
\affiliation{Oak Ridge National Laboratory, Oak Ridge, Tennessee 37831, USA}
\affiliation{JILA, University of Colorado/NIST, Boulder, CO 80309, USA}
\author{J. Garcia}%
\affiliation{Oak Ridge National Laboratory, Oak Ridge, Tennessee 37831, USA}
%\affiliation{add details}
\author{B.J. Lawrie}
\affiliation{Oak Ridge National Laboratory, Oak Ridge, Tennessee 37831, USA}\email{lawriebj@ornl.gov}

\date{\today}% It is always \today, today,
             %  but any date may be explicitly specified

\begin{abstract}
Nonlinear interferometers that replace beamsplitters in Mach-Zehnder interferometers with nonlinear amplifiers for quantum-enhanced phase measurements have drawn increasing interest in recent years, but practical quantum sensors based on nonlinear interferometry remain an outstanding challenge. Here, we demonstrate the first practical application of nonlinear interferometry by measuring the displacement of an atomic force microscope microcantilever with quantum noise reduction of up to 3~dB below the standard quantum limit, corresponding to a quantum-enhanced measurement of beam displacement of $\mathrm{1.7~fm/\sqrt{Hz}}$. Further, we show how to minimize photon backaction noise while taking advantage of quantum noise reduction by transducing the cantilever displacement signal with a weak squeezed state while using dual homodyne detection with a higher power local oscillator. This approach offers a path toward quantum-enhanced broadband, high-speed scanning probe microscopy.\footnote{This manuscript has been authored by UT-Battelle, LLC, under contract DE-AC05-00OR22725 with the US Department of Energy (DOE). The US government retains and the publisher, by accepting the article for publication, acknowledges that the US government retains a nonexclusive, paid-up, irrevocable, worldwide license to publish or reproduce the published form of this manuscript, or allow others to do so, for US government purposes. DOE will provide public access to these results of federally sponsored research in accordance with the DOE Public Access Plan (http://energy.gov/downloads/doe-public-access-plan).}

\end{abstract}

\pacs{Valid PACS appear here}

\maketitle

%\section{I. Introduction}
Over the past four decades, squeezed states of light have been developed for quantum-enhanced interferometry capable of resolving signals beyond the photon shot noise limit (SNL)~\cite{Caves,giovannetti2004quantum,lawrie2019quantum,ma2017proposal}. More recently nonlinear interferometers (NLIs), SU(1,1) interferometers in which beamsplitters are replaced with nonlinear amplifiers, have arisen as an additional application of interferometry that relies on the squeezing Hamiltonian~\cite{OuNatComm,Ou,jing2011realization,lukens2016naturally,li2016phase,du2018absolute}. In parallel with the development of quantum-enhanced interferometry, classical beam displacement measurements relying on segmented photo-detection found widespread application in sensing and microscopy~\cite{meyer1988,putman1992detailed,meyer1990simultaneous,smith1995limits}. These sensors and microscopes are frequently operated at or near the SNL. The same fundamental limit holds for beam displacement measurements based on Michelson-type interferometers where displacement is proportional to phase~\cite{putman1992detailed}. Just as squeezed light was shown to help surpass the SNL in interferometers~\cite{Caves}, it has been shown that beam displacement measurements can surpass the SNL using squeezed states of light~\cite{Fabre,treps2003quantum,pooser_ultrasensitive_2015}.

Atomic force microscopy (AFM) relying on segmented-photodiode beam displacement measurements or Michelson interferometric readout is now well understood to be limited by a variety of noise sources, including photon shot noise, laser backaction noise, laser pointing stability, mechanical vibration, electronic noise of the detector, and thermal noise resulting from the resonant cantilever interactions with the surrounding heat bath~\cite{milburn1994quantum,smith1995limits,meyer1988,fukuma2006development,labuda2012stochastic}.  Many of these noise sources can be minimized through proper system design, but the standard quantum limit (SQL), defined as the quadrature sum of backaction noise and shot noise, can only be surpassed with quantum states of light~\cite{Caves,jaekel1990quantum,giovannetti2004quantum}. Thermal noise typically exceeds the SQL at the cantilever's resonance frequency~\cite{milburn1994quantum,smith1995limits}. When off-resonance, AFMs operate in the photon shot noise limited regime where the power cannot be further increased because of thermal-effects and backaction noise~\cite{pooser_ultrasensitive_2015}.

AFMs are generally operated at the cantilever resonance frequency because resonant operation provides 1-2 orders of magnitude improvement in signal to noise ratio (SNR)~\cite{albrecht1991frequency}.  However, AFM performed on resonance effectively imposes a narrow band amplifier on the microscope, substantially narrowing the available material bandwidth that can be probed and slowing measurements due to micromechanical ringdown effects. For a wide variety of high speed microscopies, including mass sensing or protein pulling/unfolding experiments, it is preferred to operate off resonance despite the reduction in sensitivity~\cite{olcum2015high,hoffmann2003dynamics,jarvis1999off}. Sufficient reduction in the noise floor provided by a squeezed readout field would enable non-resonant AFM capable of probing broadband RF nanoscale material properties.

While quantum sensors relying on squeezed light sources are increasingly capable of surpassing the sensitivity of optimized classical sensors ~\cite{aasi2013enhanced,pooser_ultrasensitive_2015,dowran2018quantum,otterstrom_nonlinear_2014,pooser_plasmonic_2015,taylor2014subdiffraction,fan2015quantum,taylor2013biological,lawrie2013toward}, the quantum noise reduction in these sensors is highly dependent on optical loss, and the difficulty involved in controlling the spatial distribution of quantum correlations has limited the practicality of squeezed beam displacement measurements with segmented photodetectors~\cite{treps2003quantum,pooser_ultrasensitive_2015}. NLIs offer the potential to outperform classical interferometers by a factor proportional to their nonlinear gain~\cite{OuNatComm,Ou,jing2011realization}, and their benefit can be traced to the amount of squeezing generated by the nonlinear amplifier. Truncated NLIs replace the second nonlinear amplifier of a NLI with balanced homodyne detection while maintaining the phase sensitivity of the SU(1,1) interferometer~\cite{anderson2017phase,gupta2018optimized,prajapati2019polarization}. While a traditional interferometer signal is proportional to the classical sensing power, a NLI maintains the same signal scaling while reducing the noise floor by a factor proportional to the gain, thereby achieving the same sensitivity as a full NLI. In this letter, we describe a truncated NLI measurement of the displacement of an AFM microcantilever, and we show that this quantum microscope fully eliminates the previous requirement for spatial control over quantum correlations~\cite{pooser_ultrasensitive_2015,treps2003quantum}, drastically improving its practicality while allowing for classically inaccessible interferometric measurements.

\begin{figure}[t]
\centering{\includegraphics[width=0.9\columnwidth]{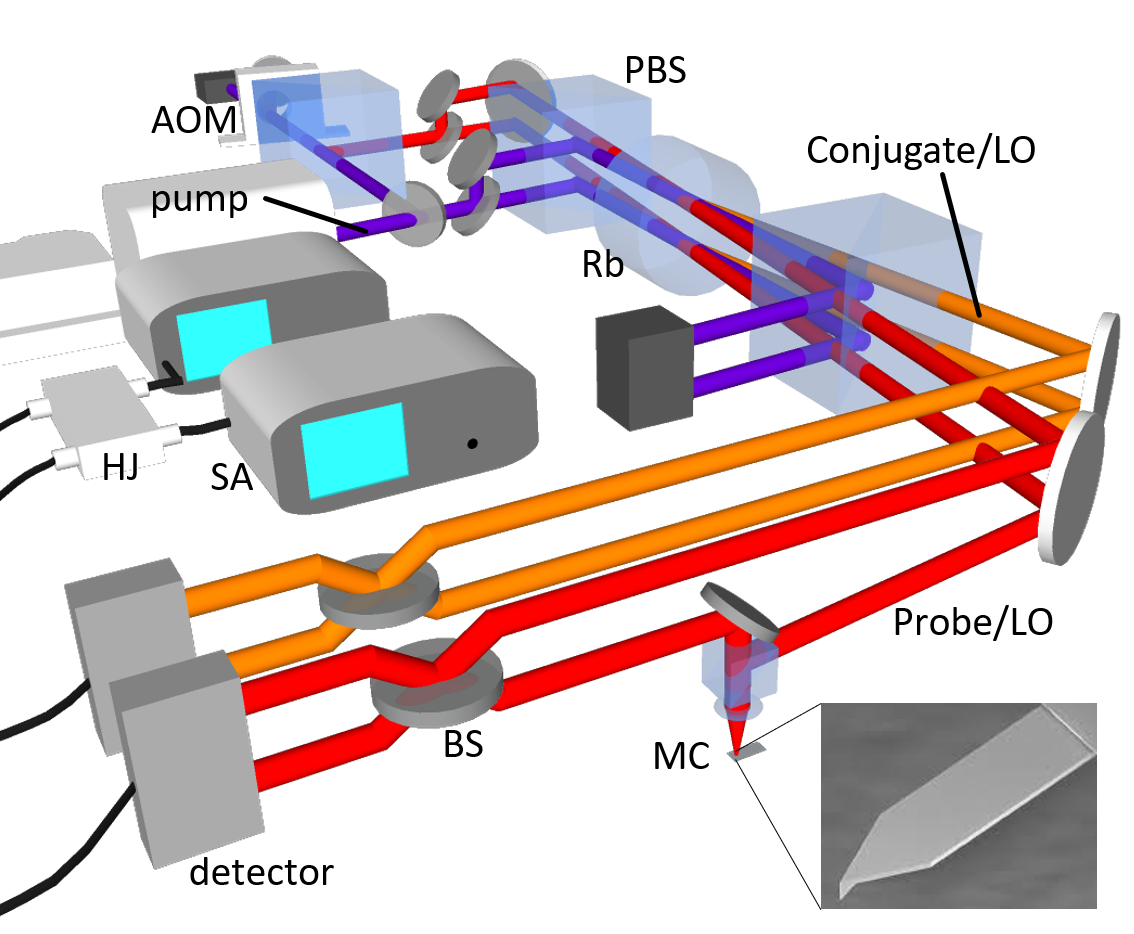}}
\caption{An acousto-optic modulator (AOM) redshifts the probe 3.042GHz from the pump, and a polarizing beamsplitter (PBS) combines the cross-polarized pump and probe in the Rb vapor cell.  Two pump beams and two variable-power probe beams generate a weak two-mode squeezed state and the corresponding high power LOs. The probe and its LO are illustrated in red, and the conjugate and its LO are illustrated in orange. Varying the relative power of the two seed probe beams enables the probe and conjugate to be easily converted into LOs and vice versa. Either the probe or the probe's LO is reflected from the AFM cantilever (MC) before the twin-beam LOs are mixed with the twin-beam squeezed states on 50/50 beam splitters (BS) in a dual homodyne detector.}

\label{fig:schematic}
\end{figure}

The truncated NLI demonstrated here relies on the same four-wave-mixing process that has previously been used for squeezed AFM cantilever beam displacement measurements on segmented photodetectors~\cite{pooser_ultrasensitive_2015}. 
As in that experiment, and as shown in Fig.~\ref{fig:schematic}, a strong pump beam and a weak probe seed beam redshifted 3.042 GHz from the pump are mixed at an angle of 0.3$^{\circ}$ in a 12.7 mm long $^{85}$Rb vapor cell held at roughly 116.5$^{\circ}$C, resulting in measured intensity difference squeezing of up to 5 dB relative to the SNL when measured directly after the vapor cell. As previously reported, a tapered amplifier operating as a master oscillator power amplifier provided a stable, compact, and low cost laser source~\cite{lawrie2016robust}. Here, the truncated NLI replaces intensity difference measurements with dual homodyne interferometry. For all of the datasets reported here, the power of the squeezed probe and conjugate fields were roughly $\mathrm{1.5 \mu W}$ and $\mathrm{1.4 \mu W}$ respectively, and the power of the probe and conjugate LOs were roughly $\mathrm{110 \mu W}$ and $\mathrm{70 \mu W}$ respectively. A proportional-integral controller was used to phase-lock the measurement at the center of the interference fringes. The spectrum analyzer settings included a $\SI{10}{\kilo \hertz}$ resolution bandwidth, $\SI{30}{\hertz}$ video bandwidth, 0.5 s sweep time, and 20 averages. 

AFM beam displacement measurements were performed with either the probe or the probe's LO reflected from a gold-coated AFM microcantilever with a fundamental resonance of 13 kHz and a force constant of 0.2~N/m in a Bruker piezo-actuated AFM mount driven at $\SI{737}{\kilo \hertz}$. Notably, using either the probe or the probe's LO to transduce the microcantilever motion results in qualitatively similar responses despite fundamentally different operating regimes. When the probe is reflected from the cantilever, the 5\% loss on the cantilever results in 0.2~dB reduction in squeezing.  In contrast, when the probe's LO is reflected from the cantilever, the only reduction in quadrature squeezing occurs as a result of reduced modematching, and that reduction can be minimized by passing the probe through the same optical train, with the cantilever replaced by a macroscopic mirror, as shown in Fig.~\ref{fig:schematic}. However, as seen in equations ~\ref{eq:SNL} and ~\ref{eq:back} below, using the probe rather than the probe LO to transduce the cantilever response virtually eliminates backaction noise from the readout, whereas a high power LO could induce backaction noise in excess of the photon shot noise if it were used to transduce the cantilever response.   

%\section{Fundamental limits}

For interferometric beam displacement measurements where the signal is purely based on the phase acquired by small displacements of the AFM cantilever, the cantilever displacement noise arising from the shot-noise-limit is
\begin{equation}
\langle \Delta (\hat{X}_-)^{2} \rangle _{SNL}=\frac{1}{4\pi^2}{\frac{h c \lambda \Delta f}{2 P_{tot}}}
\label{eq:SNL}
\end{equation}
for wavelength $\lambda$, total optical power incident on the detectors $P_{tot}$, and measurement bandwidth $\Delta f$~\cite{smith1995limits}. 

The backaction noise induced onto an optical readout field by a micromechanical cantilever operating near resonance with quality factor Q and spring constant k is
\begin{equation}
\langle \Delta (\hat{X}_-)^{2} \rangle _{back} = \frac{4Q^2}{k^2}\frac{2P h \Delta f}{c \lambda},
\label{eq:back}
\end{equation}
where $P$ is the optical power incident on the cantilever~\cite{smith1995limits}. For measurements performed off resonance, as in the data presented here, the backaction noise is significantly reduced as $Q \rightarrow 1$. The SQL is then given by the quadrature sum of equations 1 and 2. Backaction noise will still contribute to the SQL at sufficiently high powers, but the present work was conducted deep into the shot-noise limited regime. 

In the presence of squeezing, the smallest measurable beam displacement signal is proportional to the inverse square root of the intensity and the squeezing parameter, $r$~\cite{Fabre} so that equation~\ref{eq:SNL} becomes:
\begin{equation}
\langle \Delta (\hat{X}_-)^{2} \rangle _{SNL}=\frac{1}{2\pi e^r}\sqrt{\frac{h c \lambda \Delta f}{2 P_{tot}}}
    \label{SMQLBD}
\end{equation}
For zero squeezing (in a coherent laser readout for instance) one recovers the standard minimum resolvable displacement proportional to the inverse square root of intensity. Eq.~\ref{SMQLBD} applies to both absolute displacement measurements and relative displacement measurements. The only requirement is that, for $r>0$, the readout field shows relative intensity squeezing if a differencing measurement is used as the transduction mechanism.
Thus, Eq.~\ref{SMQLBD} can straightforwardly be extended to the twin beam, multimode squeezing case used in Ref.~\cite{pooser_ultrasensitive_2015} by changing the measurement to the difference in displacement between two beams of light, rather than using only a single beam of light to transduce the signal. 

For relative quadrature difference measurements of two-mode squeezed states, one can show that 
\begin{align}
\begin{split}
\left \langle \Delta (\hat{X}_-)^{2} \right \rangle &= \eta \bigg(\sinh{2r}\tanh{2r}(2\cos{(\theta_p+\theta_c-\phi}))\\
&+\cosh{2r}-\tanh^2{2r}+\sinh{2r}\tanh{2r}-1 \bigg)\\&+\tanh^2{2r}+1
\end{split}
\end{align}
where $\eta$ is the composite detection efficiency, $\theta_p$ and $\theta_c$ are the homodyne phases for the probe and conjugate, and $\phi$ is the phase shift in the probe arm of the interferometer~\cite{anderson2017phase}.

\begin{figure}[b]
\centering
\includegraphics[width=\columnwidth]{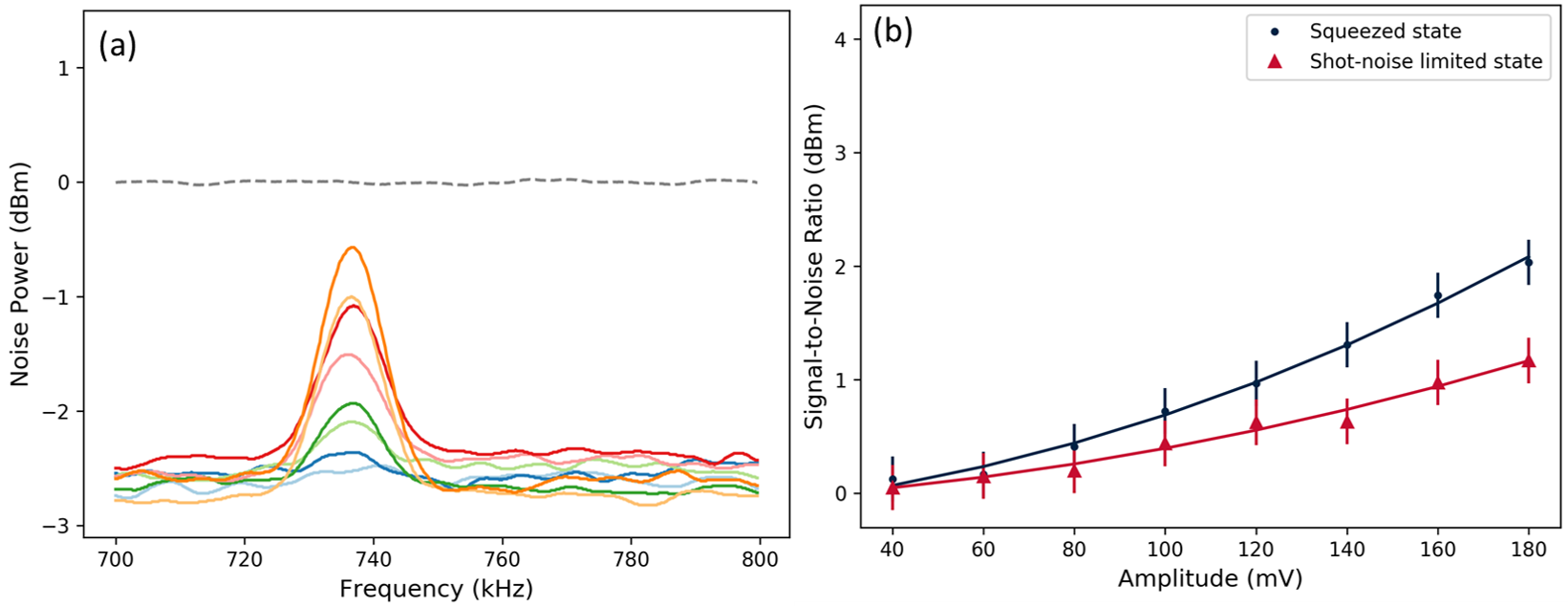}
\caption{(a) Spectrum  analyzer traces of microcantilever displacement normalized to SNL (dashed gray line) when a weak probe is reflected from the microcantilever before dual homodyne detection. (b) SNR of microcantilever displacement using squeezed light (circles) and coherent light (triangles).}
\label{fig:squeezedfig}
\end{figure}

In the ideal case where $\eta=1$ and $\theta_p = \theta_c = \pi/2$, and for minimum phase shifts of $\phi \approx 0$
\begin{equation}
  \frac{\left \langle \Delta (\hat{X}_-)^{2} \right \rangle}{SNL} = \frac{1}{2G-1},
\end{equation}\label{quadsqz}
where $G$ represents the gain in a nonlinear amplifier used to generate two-mode squeezing. This expression describes quantum noise reduction below the SNL for all $G>1$. Further it holds when the quadratures being measured are bright fields and when the phase quadrature corresponds to an optical phase, resulting in a quantum enhancement in the SNR compared with shot-noise-limited classical interferometry. Writing in terms of the squeezing parameter, one obtains qualitatively the same scaling of Eq.~\ref{SMQLBD} for the minimum resolvable relative phase measurement between to the two modes of the two-mode squeezed state.

It was also previously shown that noise in  direct intensity measurements of differential beam displacement scales as $\langle \Delta (\hat{X}_-)^{2} \rangle \propto 1/(2G-1)$~\cite{pooser_ultrasensitive_2015}. This means that a differential beam displacement measurement using direct intensity detection has an SNR scaling equivalent to a relative phase measurement. But if a nonlinear amplifier is used to produce squeezed quadratures, then such a relative phase measurement is equivalent to a phase measurement with a truncated NLI~\cite{gupta2018optimized}.

In addition, the truncated NLI has been shown in theory to have the same phase sensitivity as the full nonlinear interferometer~\cite{anderson2017phase}.  Given this equivalence between the phase sensitivity of NLIs and truncated NLIs and the equivalence between the sensitivity of NLIs and quantum-enhanced direct intensity detection, we expect a phase-measurement based truncated NLI to be capable of the same SNR and noise reduction shown in the intensity readout truncated NLI shown in~\cite{pooser_ultrasensitive_2015}.  The crucial difference for the truncated NLI is that all spatial mode dependence is contained in the mode matching between the local oscillator and the signal modes.

\begin{figure}[t]
\centering
\includegraphics[width=\columnwidth]{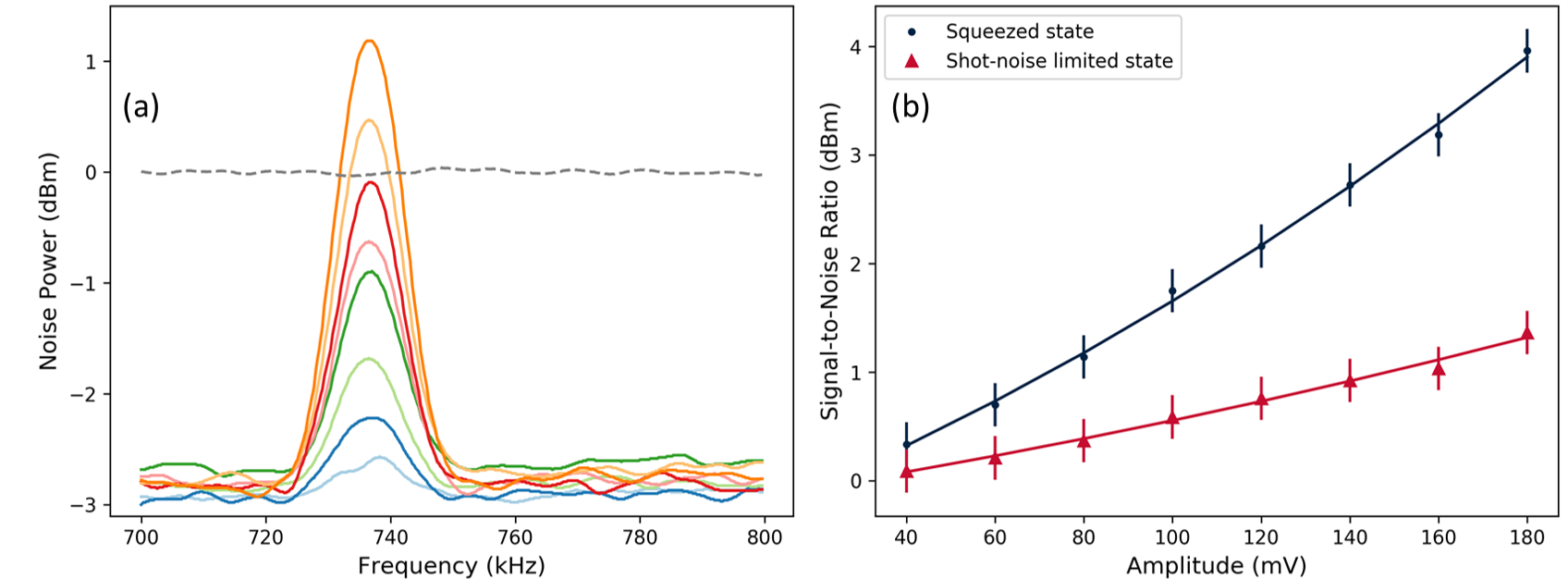}
\caption{(a) Spectrum  analyzer traces of microcantilever displacement normalized to SNL (dashed gray line) when the local oscillator is reflected from the microcantilever before dual homodyne detection. (b) SNR of microcantilever displacement using squeezed light (circles) and coherent light (triangles).}
\label{fig:LOfig}
\end{figure}

Figure~\ref{fig:squeezedfig}a illustrates the measured phase-sum signal from the dual homodyne measurement when the weak squeezed state is reflected from the AFM cantilever and the piezo actuator is driven at $\SI{40}{\milli \volt}$ to $\SI{180}{\milli \volt}$ (teal to orange respectively).  Figure~\ref{fig:squeezedfig}b illustrates the measured signal to noise ratio for each signal in Fig.~\ref{fig:squeezedfig}a along with the corresponding signal to noise ratios for shot-noise limited measurements.  The measured phase-sum squeezing varied from 2.6-2.8 dB below the SNL.  

Figures~\ref{fig:LOfig}a and b illustrate the measured phase sum signal and SNR respectively when the local oscillator is reflected from the AFM cantilever instead of the weak squeezed state. All other experimental parameters remained the same. Here, the measured squeezing varied from 2.8-3.0 dB below the SNL.  Notably, the measured SNR is almost 2dB smaller in Fig. 2 than in Fig. 3.  This suggests that some intensity difference signal, arising from relative misalignment of the probe LO, is present in these results.  However, because only 2dB difference in SNR is seen despite roughly 70x difference in optical power, we can conclude that the results seen here are mostly a relative phase measurement.

For the experimental parameters described above, the photon SNL was $\mathrm{3.3 fm/\sqrt{Hz}}$, the backaction noise when the local oscillator was incident on the cantilever was $\mathrm{243 zm/\sqrt{Hz}}$, and the backaction noise when the weak probe was incident on the cantilever was $\mathrm{29 zm/\sqrt{Hz}}$.  The experiments performed here were limited to a low gain regime by the available pump power and by Doppler broadening in the Rb vapor cell, but the LO power can plausibly be increased by two orders of magnitude by increasing the seed probe power, the pump power, and the vapor cell temperature without detrimentally effecting the dual homodyne detection. In such a regime, the measurements would still be shot noise limited if the LO were used to readout the cantilever displacement, but laser heating of the cantilever would modify the cantilever and material properties.  Thus, in the high LO power regime, the low power squeezed state must be used to readout the cantilever beam displacement.  The measured phase-sum squeezing of up to 3 dB was also limited by operation in the low gain regime, but squeezing in excess of 10 dB is possible with this squeezed light source~\cite{liu2019interference}. Further, unlike other recent demonstrations of quantum sensors where substantial loss is intrinsic to the sensor~\cite{fan2015quantum,pooser_plasmonic_2015,dowran2018quantum,otterstrom_nonlinear_2014}, only 5\% optical loss is introduced by the current sensor design. The loss can be substantially further reduced by improving the reflective coating on the AFM cantilever.  Finally, since the optimal measurement in this system is a relative phase sum measurement, one can transduce a signal onto both the probe and conjugate fields (or both LOs) and measure the phase sum while maintaining quantum noise reduction. Thus, by reflecting the probe and conjugate fields from the same cantilever, the total signal could be doubled while still taking advantage of the improved dynamic range of the previously described NLI design. 

By utilizing both the probe and conjugate fields to transduce the cantilver displacement while optimizing the LO power, the available squeezing, and the optical loss, it is therefore possible to obtain greater than two orders of magnitude further improvement in SNR compared with the measurements reported here using current technology.  Such an enhancement in sensitivity is comparable to the enhancement provided by operating an AFM at the micromechanical resonance frequency~\cite{smith1995limits}.  As a result, an optimized truncated nonlinear interferometric AFM should enable quantum-enhanced atomic force microscopy where the enhancement in SNR enabled by high power LOs and optimized quantum noise reduction would offset the advantage of resonant operation.  This form of quantum microscopy would therefore provide a broadband modality in which the full RF sideband of the measured photocurrent could be used to characterize high-speed dynamics in materials.

This research was supported by the U. S. Department of Energy, Office of Science, Basic Energy Sciences, Materials Sciences and Engineering Division. The experimental concept was conceived and initial experiments were performed as part of the Laboratory-Directed Research and Development Program of Oak Ridge National Laboratory, managed by UT-Battelle, LLC for the U.S. Department of Energy. N.S., E.B., and J.B. were supported by the U.S. Department of Energy, Office of Science, Office of Workforce Development for Teachers and Scientists (WDTS) under the Science Undergraduate Laboratory Internship program. J.G. was supported by the W.M. Keck Foundation.

\end{document}